\documentclass[aps,prd, onecolumn,notitlepage,floatfix,longbibliography,nofootinbib]{revtex4-1}
\usepackage[textwidth=15cm]{geometry}
\usepackage{amsmath}
\usepackage{amssymb}
\usepackage{amsfonts}

\usepackage{url}
\usepackage[colorlinks=true,urlcolor=blue,linkcolor=blue,citecolor=blue]{hyperref}
\usepackage{braket}

\begin{document}	

\title{Applications of the Hawking Energy on lightcones in cosmology}
\author{Dennis Stock}
\email[]{dennis.stock@zarm.uni-bremen.de}
\affiliation{University of Bremen, Center of Applied Space Technology and 
	Microgravity (ZARM), 28359 Bremen, Germany}

\begin{abstract}
The past lightcone of an observer in a cosmological spacetime is the unique geometric structure directly linked to observations. After general properties of the Hawking energy along slices of the past lightcone have previously been studied, the present work continues along this path by providing explicit cosmological applications of the Hawking energy associated with a lightcone. Firstly, it is shown that amongst all two-dimensional non-trapped spheres with equal area and average matter density, a shear-free matter distribution maximizes the Hawking energy for sufficiently high densities. Secondly, a Robertson-Walker-reference slice is constructed for every lightcone slice based on area and energy. Thirdly, after a few pedagogical examples in concrete FLRW spacetimes, the implications of monotonicity of the energy down the lightcone are explored, arriving at two new bounds on the cosmic fluid's density and equation-of-state parameter.
\end{abstract}

\maketitle


\section{Introduction}
Inhomogeneities in the Universe manifest themselves on many different scales, for instance stars, galaxies, and galaxy clusters. Yet on the largest scales, most cosmological observations are well-described by a homogeneous FLRW spacetime, despite some tensions \cite{Buchert:2015wwr,Verde:2019ivm}. Therefore, one would like to assign a smooth and homogeneous reference spacetime to an inhomogeneous universe. The question of what the best smooth background model is, and in which norm it fits best, is usually referred to as the fitting problem in inhomogeneous cosmology \cite{Ellis:1987zz,Kolb:2009rp,Clarkson:2011zq,ellis_maartens_maccallum_2012}. There are multiple ways to perform such a fitting procedure, for instance along spacelike hypersurfaces in a 3+1 decomposition. Note, however, that in this case the time evolution of the best-fit spacelike hypersurfaces is in general different from the time evolution of an FLRW universe. Alternatively, the fitting can be done along the null hypersurface of the past lightcone, avoiding the arbitrariness of a foliation choice in the 3+1 case, but instead mixing spatial and temporal information. Also, the past lightcone of the observer is the unique geometric structure directly linked to cosmological observations. In principle, one can fit either geometric quantities, for example curvatures, or direct observations such as distance-redshift relations. An active strategy for arriving at a homogeneous distribution of matter is to define an averaging procedure, by which one hopes to smooth out local inhomogeneities to arrive at the FLRW regime, see for instance \cite{Gasperini:2011us,Fanizza:2019pfp,Buchert:1999er,Buchert:2001sa,Bonvin_2015,Buchert:2002ht,Zalaletdinov:2008ts} and also the reviews \cite{Clarkson:2011zq,ellis_maartens_maccallum_2012,Buchert:2011sx} with references therein. Selecting a good background model is essential for determining the magnitude of deviations from it, which is still under debate \cite{Clarkson:2011zq,ellis_maartens_maccallum_2012,Green:2014aga,Buchert:2015iva,Larena:2008be}.

Amongst other approaches, the current work advocates the lightcone set-up because of its direct connection to observations. While studying the effect of inhomogeneities in the matter distribution, one should compare domains of equal size and total matter or energy, as it was advocated in \cite{Ellis:1998ha}. Otherwise, a variation in the total energy may also lead to different dynamics. In the present context of lightcones, a natural domain is provided by constant affine parameter slices of the lightcone, each of which comes with an associated area and Hawking energy \cite{Hawking:1968qt,Szabados:2009eka}. Phenomenologically, the Hawking quasi-local energy quantifies the lightbending of the lightcone generators. In direct continuation of \cite{Stock:2020oda}, in which its general properties for lightcones were studied, the present work provides a direct application of these concepts, comparing domains based on energy, area and matter density in section \ref{sec:comparing}. In section \ref{sec:FLRWreference}, energy and area are used to assign a Robertson-Walker (RW) reference slice to every constant affine parameter slice. RW reference refers to a homogeneous slice which is described by a constant curvature metric. As discussed in more detail in section \ref{sec:FLRWreference}, the time evolution of such a RW slice is in general different from the FLRW time evolution. After providing a few pedagogical examples of the Hawking energy in flat FLRW spacetimes in section \ref{sec:examples}, monotonicity of the Hawking energy down the lightcone is exploited in section \ref{sec:monotonicity} in order to arrive at bounds for the density and equation-of-state parameter of the cosmic fluid.


\section{Geometrical set-up}
We use the signature convention $(-+++)$ and units in which $G=c=1$. The universe is assumed to be filled with an effective, ideal fluid with normalized, geodesic 4-velocity $u^a$, i.e. $u_a u^a=-1$ and $\nabla_u u^a=0$, satisfying the Einstein field equations (EFEs)
\begin{equation}
R_{ab}-\frac{1}{2}R g_{ab}=8\pi T_{ab}\quad,
\end{equation}
with energy-momentum tensor
\begin{equation}
T_{ab}=(\rho+P)u_au_b+P g_{ab}\quad,
\label{eq:energymomentum}
\end{equation}
and equation of state (EOS)
\begin{equation}
P=w\rho
\end{equation}
with EOS parameter $w$. This effective fluid may consist of a mixture of different components, such as dust $(w=0)$, radiation $(w=1/3)$, or a cosmological constant $(w=-1)$. Additionally, the dominant energy condition (DEC), $|w|\leq 1$, is assumed to hold.

Given a fluid-comoving observer, an event $p\in M$ equipped with the unit timelike vector (4-velocity) $u^a$, we study the past lightcone $C^-(p)$ issued at $p$, i.e. the set of points in spacetime that can be reached along a past-pointing null geodesic emanating from $p$. We assume the lightcone to be caustic-free which implies that its topology is $\mathbb{R}^+\times S^2$. In fact, this is always true in a sufficiently small neighbourhood of $p$ corresponding to the injectivity domain of the exponential map generating $C^-(p)$. Physically, we operate in the weak lensing regime, where inhomogeneities may cause a deformation of the lightcone, but are sufficiently small to avoid multiple-imaging through lightcone self-intersections. The lightcone can then be sliced into two-dimensional spacelike surfaces $S_\lambda\simeq S^2$ of constant affine parameter $\lambda$. Any such slice $S_\lambda$ has two distinct null geodesic congruences, both of which are orthogonal to $S_\lambda$. Their generators are denoted by $l^a$ and $n^a$ respectively and $l^a$ is identified with the generators of $C^-(p)$. Furthermore, we decompose $l^a$ and $n^a$ into $u^a$ and a spacelike unit-vector $v^a$ orthogonal to $u^a$ and $S$, i.e. $v_a v^a=1$ and $v_a u^a=0$, and fix the normalisation of $n^a$ by demanding $n^al_a=-1$:
\begin{equation}
l^a=(u\cdot l)\left(-u^a+v^a\right)\quad\&\quad n^a=\frac{1}{2(u\cdot l)}\left(-u^a-v^a\right)\quad,
\label{eq:decomposition}
\end{equation}
where $u\cdot l=u_al^a$ is a function of the affine parameter and encodes the rescaling freedom of $l^a$. In the following, we fix the remaining freedom in the affine parameter $\lambda$ by demanding $u_al^a|_p=1$, which determines the slicing of $C^-(p)$. Then, the general redshift formula relating emitter and observer,
\begin{equation}
1+z =  \frac{(u_al^a)|_\mathrm{em}}{(u_al^a)|_\mathrm{obs}}\quad,
\end{equation}
with the observer located at $p$, yields
\begin{equation}
(u\cdot l)(\lambda)= 1+z(\lambda)\quad.
\label{eq:normalization}
\end{equation}
Redshift and affine parameter are related via the following evolution equation, cf. \cite{Ehlers:1993gf} \cite{ellis_maartens_maccallum_2012}:
\begin{equation}
\partial_\lambda z\equiv \dot{z}= (1+z)^2\left[\tilde{\sigma}_{ab}v^av^b+\frac{1}{3}\tilde{\theta}\right]\quad,
\label{eq:zdot}
\end{equation}
where $\tilde{\theta}$ \& $\tilde{\sigma}_{ab}$ denote the expansion scalar and shear tensor of the cosmic fluid. It is stressed, that these quantities can in principle be determined via observations, see e.g. the discussion in \cite{ellis_maartens_maccallum_2012}. In an FLRW universe, $\tilde{\sigma}_{ab}=0$ and $\frac{1}{3}\tilde{\theta}=H$, where $H(z)=\sqrt{\frac{8\pi}{3}\rho(z)-K(1+z)^2}$ is the Hubble function, with $K>0$, $K=0$, or $K<0$ the spatial curvature parameter with dimension $\mathrm{length}^{-2}$. Using this decomposition of $l^a$ and $n^a$ for the following components of the energy momentum tensor yields:
\begin{equation}
T_{ln}=\frac{\rho}{2}\,(1-w)\quad,\quad T_{ll}=(u\cdot l)^2\rho\,(1+w)
\quad,\quad T=\rho(3w-1)\quad.
\label{eq:Tcomponents}
\end{equation}

Of course, the affine parameter slicing adopted here is just one of many admissible ways to foliate the lightcone. It provides a monotonic distance function along the null generators, in contrast to other distance measures, such as the area distance. Additionally, it simplifies the calculations in section \ref{sec:FLRWreference}, but it is stressed that the results can be extended for other foliations.

\section{Comparing lightcone domains}
\label{sec:comparing}

In order to give a meaning to the phrase of comparing two spacetimes, one typically compares certain invariant quantities within a fixed spacelike domain. In general, there are many ways to define such a domain, however, if the domain is required to be directly related to cosmological observations, the natural choice is to take lightcone slices, because the lightcone is the unique geometric structure directly linked to cosmological observations, since all cosmologically relevant signals travel along the cone to the observer. Any of these spherical spacelike lightcone slices $S$ comes with an associated area
\begin{equation}
A(S):= \int_S \mathrm{d}S\quad,
\end{equation}
defined as the integral of the pullback of the spacetime volume-form onto $S$. Additionally, we may wish to assign an energy to $S$. In the present context, a good measure for the energy associated with a lightcone slice is the Hawking energy \cite{Hawking:1968qt,Szabados:2009eka,Stock:2020oda}, because it relates the energy enclosed by a sphere to the amount of light bending on it, once a foliation choice is made. Therefore, using the Hawking energy provides a way to quantify the deformation of the lightcone in terms of the energy/matter content enclosed by it. Formally, for a given spacelike 2-sphere $S$, the Hawking energy is defined as
\begin{align}
E(S) &= \frac{\sqrt{A(S)}}{(4\pi)^{3/2}} \left(2\pi+\frac{1}{4}\int_S \theta_+\theta_-\,\mathrm{d}S\right)\label{eq:origdef}\\
&=\frac{\sqrt{A(S)}}{(4\pi)^{3/2}} \int_{S} \left(4\pi T_{ln}+\frac{2}{3}\pi T+\frac{1}{2}\sigma^{ab}_+\sigma_{ab}^-\right)\,\mathrm{d}S\nonumber\\
&\overset{(\ref{eq:Tcomponents})}{=}\frac{\sqrt{A(S)}}{(4\pi)^{3/2}} \int_{S} \left(\frac{4\pi}{3}\rho+\frac{1}{2}\sigma^{ab}_+\sigma_{ab}^-\right)\,\mathrm{d}S\quad,
\label{eq:energy}
\end{align}
with expansion scalars $\theta_\pm$ and shear tensors $\sigma_{ab}^\pm$ of the two null congruences orthogonal to $S$, represented by $l^a$ and $n^a$. One arrives at (\ref{eq:energy}) from (\ref{eq:origdef}) by using the Gauss-Bonnet theorem for a 2-sphere, $\int_S {}^2R\,\mathrm{d}S=8\pi$, together with the contracted Gauss equation \cite{Gourgoulhon:2005ng,Stock:2020oda},
\begin{equation}
{}^2R = 16\pi T_{ln}+\frac{8\pi}{3}T-\theta_+\theta_- + 2\sigma_{ab}^+ \sigma_-^{ab}\quad,
\label{eq:Gauss}
\end{equation}
where ${}^2R$ denotes the Ricci scalar of $S$. When comparing with other energy notions in General Relativity, see for instance \cite{Szabados:2009eka}, the Hawking energy is straightforward to compute and directly linked to light propagation via the expansion scalars. As can be seen from (\ref{eq:energy}), it not only takes into account matter contribution given by the energy-momentum tensor components, but also has a purely gravitational contribution via the Weyl tensor, entering through the shear terms, cf. (\ref{eq:shearevolution}). In particular, (\ref{eq:energy}) shows that the energy is independent of the fluid's pressure.\\

In general, an inhomogeneous domain\footnote{In the following, a domain refers to the interior of $S$ in a particular foliation. A domain is said to be homogeneous (respectively inhomogeneous), if the matter distribution inside $S$ on a given spacelike hypersurface is homogeneous (respectively inhomogeneous). Even without specifying a spacetime foliation, one may still call $S$ a homogeneous domain by referring to homogeneity on $S$, which corresponds to isotropy for the observer.} differs from a homogeneous one by the presence of the shear term $\sigma_{ab}^+\sigma_-^{ab}$ in (\ref{eq:energy}). The presence of inhomogeneities sources a non-vanishing Weyl tensor which generates shear of a null congruence via the shear evolution equation \cite{Wald:1984rg}
\begin{equation}
\dot{\sigma}_{ab}=-\theta\sigma_{ab}-C_{acbd}l^cl^d\quad.
\label{eq:shearevolution}
\end{equation}
FLRW spacetimes and other conformally flat spacetimes, in which the Weyl tensor vanishes identically, are of Petrov-type O, and hence, an initially shear-free null congruence remains shear-free: $\sigma_{ab}=0$. This holds especially for the null generators of a lightcone, which are shear-free sufficiently close to the vertex point $p$.

In fact, general bounds on the shear contribution can be derived in the following way. Integrating the contracted Gauss equation (\ref{eq:Gauss}) over $S$ and dividing by $A(S)$, yields
\begin{equation}
\braket{\sigma_{ab}^+\sigma_-^{ab}}_S= \frac{4\pi}{A}+\frac{1}{2}\braket{\theta_+\theta_-}_S-\frac{8\pi}{3}\braket{\rho}_S \quad,
\end{equation}
where $\braket{f}_S=A^{-1} \int_S f\,\mathrm{d}S$ denotes the average of $f$ over $S$. By assumption, $\theta_+\theta_-\leq 0$ in the weak lensing regime, and thus, the second and third term is negative. Upon exploiting the positivity of $E$ in the weak lensing regime \cite{Stock:2020oda} in the definition (\ref{eq:origdef}), we arrive at a bound of the shear-term in terms of the average density:
\begin{equation}
-\frac{8\pi}{3}\braket{\rho}_S\leq \braket{\sigma_{ab}^+\sigma_-^{ab}}_S \leq \frac{4\pi}{A}-\frac{8\pi}{3}\braket{\rho}_S\quad.
\end{equation}
In vacuum ($\rho=0$), the bound reads $0\leq \braket{\sigma_{ab}^+\sigma_-^{ab}}_S \leq \frac{4\pi}{A}$, implying that the shear-term is non-negative, however, as soon as $\int_S \rho\neq 0$, the shear contribution may be negative. In particular, for densities above the threshold density $\rho_*:=\frac{3}{2}\frac{1}{A}$, the shear-term $\braket{\sigma_{ab}^+\sigma_-^{ab}}_S$ turns negative. At the threshold density $\braket{\sigma_{ab}^+\sigma_-^{ab}}_S=\frac{1}{2}\braket{\theta_+\theta_-}_S$, hence, for densities larger than $\rho_*$, the average expansion term dominates over the average shear term.

As a consequence, when comparing an inhomogeneous and a homogeneous domain of equal size $A$ and average density $\braket{\rho}_S$, the difference in their Hawking energies is proportional to the shear-term $\braket{\sigma_{ab}^+\sigma_-^{ab}}_S$. As the average density is increased beyond the threshold density $\rho_*$, the energy of the inhomogeneous domain eventually becomes smaller than the energy of the respective homogeneous domain. In other words, a homogeneous distribution of matter maximizes the Hawking energy as long as $\rho\geq \rho_*$. It is stressed that this effect is only due to a different distribution of the matter, since the overall density is fixed.

\section{Constructing RW reference slices}
\label{sec:FLRWreference}
In order to interpret given data on an inhomogeneous lightcone in terms of an FLRW model, as it is often done within the standard framework of interpreting cosmological observables by treating deviations from an FLRW background as statistical fluctuations, we may wish to phrase the (inhomogeneous) data in terms of a RW metric template. Motivated by the direct link between the lightcone deformation and the Hawking energy, this can be achieved by assigning a RW reference slice $\bar{S}_\lambda$ to a lightcone slice $S_\lambda$ in an inhomogeneous spacetime by fixing the area and energy. Leaving the energy constant guarantees that occuring effects are only due to a redistribution of matter rather than a variation of the total energy. Moreover, if this procedure is applied to the whole lightcone, we arrive at a 1-parameter family of RW reference domains $(\bar{S}_\lambda)_{\lambda\geq 0}$, labelled by the affine parameter. How area, energy and matter content are related is studied in the subsequent two paragraphs, before discussing the explicit construction of the homogeneous RW bias.

\subsection{Link between area and matter content}
The area $A(\lambda)$ of the lightcone slice $S_\lambda$ is linked to the Ricci tensor via \cite{Hawking:1973uf}
\begin{equation}
\dot{\theta}_+=-\frac{1}{2}\theta_+^2-R_{ll}-\sigma_{ab}^+\sigma_+^{ab}\quad,
\end{equation}
where the dot denotes the derivative with respect to the affine parameter of the null generators $l^a$, normalized as discussed above, cf. (\ref{eq:normalization}). After using $\dot{A}(\lambda)=\int_{S_\lambda}\theta_+\,dS_\lambda$, the EFEs, and (\ref{eq:Tcomponents}), we find
\begin{equation}
\ddot{A}(\lambda)=\int_{S_\lambda}\left[\frac{1}{2}\theta_+^2-8\pi(1+z)^2 \rho(1+w)-\sigma_{ab}^+\sigma_+^{ab}\right]dS_\lambda\quad.
\label{eq:aev}
\end{equation}
In the FLRW case, the shear vanishes as mentioned above and slices of constant $\lambda$ are also surfaces of constant $z$, and fluid-orthogonal. Hence, $\theta_+$ is constant on each $S_\lambda$ and related to the area by $\dot{A}=\int_{S_\lambda}\theta_+\;\mathrm{d}S_\lambda= \theta_+ A$ . Alternatively, it can also be expressed in terms of the area distance $D$, defined by $A(S)=:4\pi D^2$:
\begin{equation}
\theta_+= \frac{\dot{A}}{A}= 2\frac{\dot{D}}{D}\quad.
\end{equation}
Using the chain rule $\dot{f}(z(\lambda))=f' \dot{z}$ and (\ref{eq:zdot}), the redshift dependence of $\theta_+$ in a FLRW spacetime may be expressed as
\begin{equation}
\theta_+(z)=2\frac{D'}{D}\dot{z}\overset{\mathrm{FLRW}}{=} 2\frac{D'}{D} (1+z)^2 H(z)\quad.
\end{equation}

Therefore, the area evolution equation (\ref{eq:aev}) in an FLRW universe with foliation as given above reads
\begin{equation}
\ddot{A}\overset{\text{FLRW}}{=}A\left[\frac{1}{2}\frac{\dot{A}^2}{A^2}-8\pi \rho(1+w)(1+z)^2\right]\quad.
\label{eq:AFLRW}
\end{equation}
Equivalently, the area distance must satisfy
\begin{equation}
\ddot{D}=-4\pi \rho(1+w)(1+z)^2 D\quad.
\label{eq:Deq}
\end{equation}
The differential equation (\ref{eq:AFLRW}) links the area to the matter content in an exact way, in contrast to the expansions of the area about Minkowski space, for instance about the lightcone vertex or within the causal diamond construction as discussed in \cite{Gibbons:2007nm,Wang:2019zhr}. However, these expansions are recovered from (\ref{eq:AFLRW}) by solving it in a series expansion in the respective regime. For a cosmological constant alone ($w=-1$), it reduces to a second order autonomous ordinary differential equation, which can be solved explicitly by taking the correct small sphere limit into account to yield the expected result $A(\lambda)=4\pi\lambda^2$. 

\subsection{Link between Hawking Energy, area and matter}
Recalling (\ref{eq:energy}), the Hawking energy is not only sensitive to the amount of matter, but also to its distribution via the shear contribution. As discussed before, inhomogeneities account for the shear term via the Weyl tensor. Two further points are emphasized here. Firstly, by varying the domain size $A$ and matter content in the integral in a suitable manner, we may easily reproduce the same energy. For instance, starting with a given domain, we can always find an arbitrarily large domain of the same energy by simply scaling down the matter part, given by the integral expression in (\ref{eq:energy}) accordingly. 
Secondly, even if the area is additionally fixed, a degeneracy between shear and matter contribution in the integral of (\ref{eq:energy}) remains. For example, when comparing an inhomogeneous domain with a shear-free one of the same size and energy, the absent shear contribution has to be compensated by an increase of $\int_{S_\lambda}\frac{4\pi}{3}\rho \,dS_\lambda$. In the FLRW case, the shear term vanishes, and after using (\ref{eq:Tcomponents}) we can perform the integral in (\ref{eq:energy}), since the integrand only depends on the affine parameter, yielding the simple expression
\begin{equation}
E(\lambda)\overset{\text{FLRW}}{=} \frac{4}{3}\pi D^3 \rho \quad.
\label{eq:EFLRW}
\end{equation}


\subsection{RW reference slices}
After having discussed how area, Hawking energy and matter are connected, we proceed with the explicit construction of reference slices. As mentioned above, we can assign a 1-parameter family of RW reference slices to the (inhomogeneous) lightcone by keeping area and energy constant. Notation-wise, we denote the RW quantities with an overbar, and choose the lightcone slicing (\ref{eq:normalization}) in what follows, but it is stressed that the resulting relations hold for other choices as well. Each RW domain may be thought of as a product of an (unspecified) averaging procedure while keeping domain size $A(z)$ and energy $E(z)$ fixed.

Note that a cosmological constant $\Lambda$ has $w=-1$, and therefore does not affect the area evolution, but still contributes to the energy. Thus, when studying lightcone areas exclusively, one cannot make a statement regarding the presence or absence of a cosmological constant, rather, the evolution of the area is sensitive to all matter \textit{except} a cosmological constant. According to (\ref{eq:aev}), any matter with $w\neq-1$ neccessarily causes the lightcone areas to be smaller than in a spacetime with only a cosmological constant present. This result was formally proven in \cite{ChoquetBruhat:2009fy}. 

The condition of equal energy $E(S_\lambda)=E(\bar{S}_\lambda)$ yields
\begin{equation}
\bar{\rho}(\lambda)=\braket{\rho}_{S_\lambda}+\frac{3}{8\pi}\braket{\sigma_{ab}^+\sigma^{ab}_-}_{S_\lambda}\quad.
\label{eq:bar1}
\end{equation}
Thus, the matter density in the RW reference slice is given by the average density plus a correction due to shear sourced by inhomogeneities. As discussed above, the shear term $\braket{\sigma_{ab}^+\sigma^{ab}_-}_{S_\lambda}$ has no definite sign in general, thus, it may lead either to an increase or decrease in the matter density. Demanding that the area $A(S_\lambda)=A(\bar{S}_\lambda)$ and its first and second derivative, $\dot{A}$ and $\ddot{A}$, must be equal yields
\begin{align}
\dot{A}(\lambda)&=\int_{S_\lambda} \theta_+\,\mathrm{d}S_\lambda\overset{\text{RW}}{=}\bar{\theta}_+\cdot A(\lambda)\quad,\\
\ddot{A}(\lambda)&=\int_{S_\lambda}\left[\frac{1}{2}\theta_+^2-8\pi T_{ll}-\sigma_{ab}^+\sigma_+^{ab}\right]\mathrm{d}S_\lambda
\overset{\text{RW}}{=}\frac{1}{2}\frac{\dot{A}^2}{A}-8\pi A\, \bar{T}_{ll}\quad.
\end{align}
Combining both yields
\begin{equation}
\bar{\rho}\left(1+\bar{w}\right)(1+\bar{z})^2=\braket{\rho(1+w)(1+z)^2}_{S_\lambda}+\frac{1}{8\pi}
\braket{\sigma_{ab}^+\sigma_+^{ab}}_{S_\lambda}+\frac{1}{16\pi}\left(\braket{\theta_+}_{S_\lambda}^2-\braket{\theta_+^2}_{S_\lambda}\right)\quad.
\label{eq:bar2}
\end{equation}
Again, the RW reference value $\bar{\rho}\left(1+\bar{w}\right)(1+\bar{z})^2$ is given by the average over the inhomogeneous lightcone slice plus corrections of definite sign. The shear contribution is non-negative, whereas fluctuations in the expansion parameter away from its average contribute negatively because the term involving the expansion scalar is proportional to minus its variance. Recall that $\bar{z}$ in an FLRW universe is a function of $\bar{\rho}$ once the spatial curvature is specified, cf. (\ref{eq:zdot}). Therefore, equations (\ref{eq:bar1}) and (\ref{eq:bar2}) together uniquely determine the cosmic fluid in each RW reference slice in terms of the respective quantities in a general and inhomogeneous universe. 

Note however, that a RW reference is only assigned slice-wise and \textit{not} for the whole lightcone. In other words, the pair $(\bar{\rho},\bar{w})$ does \textit{not} describe an FLRW fluid evolving with redshift or an affine parameter. The reason is that due to the covariant conservation of the energy-momentum tensor, FLRW spacetimes are subject to an additional conservation law, which can be written as \cite{ellis_maartens_maccallum_2012}
\begin{equation}
\partial_t\rho+3H\rho (1+w)=0\quad,
\label{eq:conservation}
\end{equation}
with cosmic time $t$ and Hubble function $H$. Using $a=(1+z)^{-1}$ and allowing the EOS parameter $w(z)$ to vary with redshift, this equation can be integrated:
\begin{equation}
\rho(z)=\rho_0\exp\left[3\int_0^z \frac{1+w(z')}{1+z'}\mathrm{d}z'\right]\quad.
\label{eq:rhoFLRW}
\end{equation}
Generically, the functional dependence of $\rho(z)$ and $w(z)$ in an inhomogeneous spacetime is different from (\ref{eq:rhoFLRW}). Therefore, the above procedure does not address the fitting problem on lightcones as described in the introduction, that is, which FLRW parameters provide the best fit to data on the whole lightcone. It rather constructs a homogeneous RW fluid for every lightcone slice individually, based on its area and energy. This slice-wise fit to a RW template metric may be interpreted as a homogeneous RW bias of the data.\\

Yet, for given area $A(z)$ or energy $E(z)$ alone, it is possible to construct an FLRW fluid for the whole lightcone as follows. Equation (\ref{eq:zdot}) can be used to express the derivative with respect to $\lambda$ in terms of $z$: $\partial_\lambda= (1+z)^2 H(z) \partial_z$. Applying this relation to the left hand side of (\ref{eq:Deq}) reads
\begin{equation}
\ddot{D}=(1+z)^2H\left[(1+z)^2HD'\right]' \quad.
\end{equation}
Additionally, (\ref{eq:rhoFLRW}) can be exploited to express $w(z)$ in terms of the density and its derivative, $1+w(z)=\frac{1}{3}(1+z)\frac{\rho'}{\rho}$. Inserting both relations into (\ref{eq:Deq}) yields
\begin{equation}
(1+z)^2H\left[(1+z)^2HD'\right]' = -\frac{4\pi}{3}(1+z)^3\rho' D\quad.
\label{eq:deqflrw}
\end{equation}
Given $D(z)$ and its derivatives, and recalling $H(z)=\sqrt{\frac{8\pi}{3}\rho(z)-K(1+z)^2}$, (\ref{eq:deqflrw}) constitues a first order differential equation for $\rho(z)$, once the spatial curvature is specified. $w(z)$ is then determined by (\ref{eq:rhoFLRW}). One could proceed to study deviations in the Hawking energy of an inhomogeneous lightcone from its FLRW counterpart of the same size.

Contrarily, if the energy is fixed instead of the area, the density can be expressed in terms of energy and distance via (\ref{eq:EFLRW}): $\rho=\frac{3}{4\pi}\frac{E}{D^3}$. Inserting into (\ref{eq:deqflrw}) results in a second order differential equation for $D(z)$. Solving for $D(z)$ makes it possible to compare the distance function in an inhomogeneous universe with its FLRW reference of equal energy. These studies are left for future work.
\section{The energy of lightcones in spatially flat FLRW spacetimes}
\label{sec:examples}
For the remainder of this work, we consider exact FLRW spacetimes. Before deriving lower bounds on the density and EOS parameter in the next section, we would like to offer a few pedagogical examples for how to explicitly compute the Hawking energy in some concrete and physically relevant spacetimes. The spacetimes under consideration here are spatially flat FLRW spacetimes ($K=0$) obeying the DEC ($|w|\leq 1$). In spherically symmetric spacetimes, the Hawking energy coincides with the Misner-Sharp energy, see for instance \cite{Szabados:2009eka}. Recall the expressions for energy (\ref{eq:EFLRW}), density (\ref{eq:rhoFLRW}), and that the area distance in FLRW spacetimes is identical to the angular diameter distance
\begin{align}
D(z)= \frac{1}{1+z}\; f\left(\int_0^z \frac{\mathrm{d}z'}{H(z')}\right)\quad\text{with}\quad f(r)=\begin{cases}
\frac{\sin (\sqrt{K}r)}{\sqrt{K}} \quad &\text{for}\; K>0\\
r \quad &\text{for}\; K=0\\
\frac{\sinh (\sqrt{-K}r)}{\sqrt{-K}} \quad &\text{for}\; K<0\quad,
\end{cases}\label{eq:Dflrw}
\end{align}
with $H^2(z)=\frac{8\pi}{3}\rho(z)-K(1+z)^2$. In the following, we study the Hawking energy (i) for spacetimes in which $w=$const., (ii) for a fluid with redshift-dependent EOS-parameter $w(z)\geq 0$, and (iii) in a dust universe with a positive cosmological constant.

\subsection{Fluids with $w$= const.}
For constant $w$, (\ref{eq:rhoFLRW}) can be integrated to yield $\rho(z)=\rho_0(1+z)^{3(1+w)}$. Inserting $\rho$ and $D$ into (\ref{eq:EFLRW}), the energy reads
\begin{equation}
E(z)= \frac{1}{2}\sqrt{\frac{3}{8\pi\rho_0}}\left(\frac{2}{1+3w}\right)^3  (1+z)^{3w}\left[1-(1+z)^{-\frac{1}{2}(1+3w)}\right]^3\quad.
\label{eq:Ewconst}
\end{equation}
Now, the Big Bang limit $z\rightarrow\infty$ for different $w$=const. is considered.
\begin{itemize}
	\item[(a)] $w>0$: For positive $w$, $\lim\limits_{z\rightarrow \infty}E(z)=\infty$ monotonically. This case includes radiation $w=\frac{1}{3}$. Despite the fact that beyond its turnaround redshift the angular diameter distance is decreasing, the density increases fast enough for the energy to diverge as $z\rightarrow\infty$.
	\item[(b)] $w=0$: For dust, $\lim\limits_{z\rightarrow \infty}E(z)=\sqrt{\frac{6}{\pi\rho_0}}$ monotonically, which is finite.
	\item[(c)] $0>w>-1$: In this case $\lim\limits_{z\rightarrow \infty}E(z)=0$, and $E(z)$ displays a maximum, due to the positivity of the energy for small spheres \cite{Horowitz:1982}.
	\item[(d)] $w=-1$: For a cosmological constant $\Lambda$ alone, $\lim\limits_{z\rightarrow \infty}E(z)=\sqrt{\frac{3}{4\Lambda}}$, which is finite.
\end{itemize}
Summarizing, for $w\geq 0$ and $w=-1$ the Hawking energy is monotonic up to the Big Bang, and positive for all cases of $w$. Generally speaking, there are two competing effects: the decreasing angular diameter distance beyond its turnaround, and the increasing matter density. Depending on which effect dominates in the limit $z\rightarrow\infty$, the energy is either infinite, finite but non-zero, or vanishing.

\subsection{Fluids with $w(z)\geq 0$}
For a general fluid with a varying equation of state $w(z)\geq 0$, (\ref{eq:rhoFLRW}) implies $\rho'=3\rho\frac{1+w(z)}{1+z}\geq 0$ because of the DEC. Furthermore, $H^2(z)=\frac{8\pi}{3}\rho(z)$, hence $H(z)>0$. Then, calculating the derivative of $E$ reads
\begin{align}
E(z)&= \frac{4\pi}{3}\rho(z)\frac{1}{(1+z)^3}\left(\int_0^z \frac{\mathrm{d}z'}{H(z')}\right)^3\qquad \Rightarrow\\
E'(z) &= \frac{4\pi\rho(z)}{(1+z)^3} \left(\int_0^z \frac{\mathrm{d}z'}{H(z')}\right)^2  \left\{\frac{w(z)}{1+z}\left(\int_0^z \frac{\mathrm{d}z'}{H(z')}\right) +\frac{1}{H(z)} \right\}\quad,\label{eq:E'FLRW}
\end{align}
which implies $E'(z)\geq 0$ since all appearing terms are non-negative. For example, this includes monotonicity for a mixture of dust and radiation. Depending on the asymptotic behaviour of $w(z)$ as $z\rightarrow\infty$, $E$ is finite or unbounded.

\subsection{Dust with positive cosmological constant $\Lambda$}
An FLRW spacetime filled with dust $w=0$ and a positive cosmological constant $\Lambda$ is a good approximation to the current cosmic era of accelerated expansion. As it is shown below, $\lim\limits_{z\rightarrow \infty}E(z)=\mathrm{const.}$  monotonically. Since densities and pressures of multiple fluid components are additive, we find for the combined EOS-parameter 
\begin{equation}
w(z)=-\frac{1}{1+\frac{\Omega_{m0}}{\Omega_{\Lambda 0}}(1+z)^3}\quad,\label{eq:w1}
\end{equation}
which starts at a negative value given in terms of the ratio of the energy densities of matter and the cosmological constant today, and $\lim\limits_{z\rightarrow\infty}w(z)=0$. For monotonicity, the curly bracket in (\ref{eq:E'FLRW}) must be non-negative. Exploiting monotonicity of the Hubble function $H(z)$ and (\ref{eq:w1}), we can derive the following lower bound:
\begin{align}
\frac{w(z)}{1+z}&\int_0^z \frac{\mathrm{d}z'}{H(z')}+\frac{1}{H(z)}\geq \frac{-1}{(1+z)\left[1+\frac{\Omega_{m0}}{\Omega_{\Lambda 0}}(1+z)^3\right]}\frac{z}{H_0}+\frac{1}{H(z)} \\
&=\frac{1}{H_0}\left(\frac{1}{\sqrt{\Omega_{\Lambda 0}}}\frac{1}{\sqrt{\frac{\Omega_{m0}}{\Omega_{\Lambda 0}}(1+z)^3+1}}-\frac{z}{(1+z)\left[\frac{\Omega_{m0}}{\Omega_{\Lambda 0}}(1+z)^3+1\right]}\right)\quad.\label{eq:rt}
\end{align}
It can be verified numerically, that (\ref{eq:rt}) is indeed positive for all $z$, provided that $\Omega_{m0}$ and $\Omega_{\Lambda 0}$ take realistic values, for example $\Omega_{m0}=\frac{1}{4}$ and $\Omega_{\Lambda 0}=\frac{3}{4}$, or for the respective values measured by the Planck mission \cite{Aghanim:2018eyx}. Therefore, $E(z)$ behaves monotonically for all redshifts.

\section{Bounds on FLRW fluids}
\label{sec:monotonicity}
Since in this section the monotonicity of the Hawking energy for caustic-free lightcones \cite{Stock:2020oda} is exploited to arrive at lower bounds for the density $\rho(z)$ and EOS parameter $w(z)$, the previous section offers examples under which conditions to expect monotonicity. In fact, assuming the DEC translates into $|w|\leq 1$, and positivity of the Hawking energy implies $\rho(z)\geq0$. According to (\ref{eq:EFLRW}), the energy in an FLRW spacetime monotonically increases with redshift as long as
\begin{equation}
E'(z)=\frac{4}{3}\pi \left(\rho' D^3+3\rho D^2 D'\right)\geq 0\quad\Leftrightarrow\quad
\frac{\rho'}{\rho}\geq -3\frac{D'}{D}\quad.
\label{eq:E'}
\end{equation}
This inequality relates the logarithmic derivative of the fluid's density to the respective value of the angular diameter distance. For the corresponding expression in a general inhomogeneous spacetime, see equation (14) in \cite{Stock:2020oda}. The functional dependence of density and EOS-parameter is given by (\ref{eq:rhoFLRW}). The DEC implies $\rho'(z)\geq 0$, hence the bound (\ref{eq:E'}) is trivially met as long as $D'(z)\geq 0$. Since the angular diameter distance $D(z)$ typically has a maximum due to the focussing property of matter (\ref{eq:Deq}), eventually $D'(z)< 0$. In other words, the bound becomes non-trivial beyond the turnaround of $D(z)$ at $z_*$. In this regime, we can derive the following two bounds on $\rho(z)$ and $w(z)$. Monotonicity of $E(z)$ implies $E(z_1)\leq E(z_2)$ for $z_1\leq z_2$ and thus
\begin{equation}
\rho(z_2)\geq \rho(z_1)\left(\frac{D(z_1)}{D(z_2)}\right)^3\quad,
\label{eq:ineq1}
\end{equation}
which again is trivially satisfied if the distance is an increasing function of redshift. Equating (\ref{eq:origdef}) and (\ref{eq:EFLRW}) at the turnaround $z_*$ of $D$, i.e. $\theta_+=0$, yields a direct relation between the density $\rho_*=\rho(z_*)$ and angular diameter distance $D_*=D(z_*)$ at turnaround:
\begin{equation}
\rho_*=\frac{3}{8\pi}\frac{1}{D^2_*}\quad.
\end{equation}
Inserting this into (\ref{eq:ineq1}) yields a lower bound on $\rho(z)$ for $z>z_*$:
\begin{equation}
\rho(z)> \frac{3}{8\pi}\frac{D_*}{D^3(z)}\quad.
\label{eq:rhobound}
\end{equation}
Furthermore, using (\ref{eq:rhoFLRW}) to express $\frac{\rho'}{\rho}$ in terms of $w$ yields the following lower bound on $w$ via (\ref{eq:E'}):
\begin{equation}
2\overset{\text{(DEC)}}{\geq}1+w(z)\geq -(1+z)\frac{D'}{D}\overset{\text{(DEC)}}{\geq} 0\quad.
\label{eq:wbound}
\end{equation}
It is stressed that both bounds are given in terms of observables, namely, the redshift and the area distance together with its first derivative.


\section{Discussion \& conclusions}
The current work provides a follow-up on the general results in \cite{Stock:2020oda} about the Hawking energy of lightcones in cosmological spacetimes by providing three explicit applications. Firstly, the Hawking energy of a non-trapped, i.e. $\theta_+\theta_-\leq 0$, but otherwise arbitrary domain was studied while keeping its area and the average density constant. For densities above the threshold density $\rho_*=3/2 A^{-1}$, a shear-free domain maximizes the Hawking energy. Secondly, a general method was proposed to assign a RW reference to each slice of an inhomogeneous lightcone. Keeping both the energy of the slice and its area together with the first and second derivative fixed, determines the fluid in the respective RW slice uniquely. The condition that area and energy of respective constant affine parameter slices in both spacetimes should be equal corresponds to the physical requirement of comparing domains of equal size and energy. As a consequence, any variation is only due to a different distribution of matter rather than a variation of total energy. The presence of inhomogeneities is reflected in the appearance of shear terms in (\ref{eq:bar1}) and (\ref{eq:bar2}). It is stressed again that, due to the conservation law (\ref{eq:conservation}), the 1-parameter family of densities and EOS parameters $(\rho(z),w(z))$ generated by the slice-wise fitting procedure does \textit{not} describe the evolution of an FLRW fluid with redshift in general, but rather reinterprets the given inhomogeneous geometry with a RW bias. We also briefly commented on how in principle energy \textit{or} area could be used to construct an FLRW lightcone and thereby address the fitting problem. Thirdly, the Hawking energy was shown to be monotonic up to the Big Bang in a number of spatially flat FLRW spacetimes. Depending on the detailed behaviour of angular diameter distance and density, the energy may be infinite, finite, or zero in the Big Bang limit. Furthermore, the monotonicity of the Hawking energy along the lightcone can be exploited to derive bounds (\ref{eq:rhobound}) \& (\ref{eq:wbound}) on $\rho(z)$ and $w(z)$ in FLRW spacetimes. These bounds are particularly relevant for redshifts beyond the turnaround of the angular diameter distance. It would be interesting to apply these bounds directly to observational data sets. 

The discussion here was based on the foliation provided by the unique affine parameter induced by the observer's 4-velocity. One advantage is the monotonic behaviour of this parameter along the lightcone generators. However, the results can be transferred to any other slicing function, as long as they are single-valued functions of the affine parameter. Furthermore, we operated in the weak lensing regime because the presence of caustics typically introduces points conjugate to the vertex along the null generators at which $\theta_+\rightarrow-\infty$. As a consequence, the Hawking energy fails to be monotonic in general \cite{Stock:2020oda}. Since multiple imaging does occur in the Universe, it would be desirable to admit at least certain classes of caustics in the above discussion.

\begin{acknowledgements}
	The author is very grateful for the fruitful discussions with and comments of T. Buchert, D. Giulini, A. Heinesen, and Pierre Mourier. The author is also grateful to an anonymous referee for pointing out an error in an earlier version. This work is supported by the DFG Research Training Group ``Models of Gravity".
\end{acknowledgements}

\bibliography{LetterFitting}

\begin{thebibliography}{30}%
\makeatletter
\providecommand \@ifxundefined [1]{%
 \@ifx{#1\undefined}
}%
\providecommand \@ifnum [1]{%
 \ifnum #1\expandafter \@firstoftwo
 \else \expandafter \@secondoftwo
 \fi
}%
\providecommand \@ifx [1]{%
 \ifx #1\expandafter \@firstoftwo
 \else \expandafter \@secondoftwo
 \fi
}%
\providecommand \natexlab [1]{#1}%
\providecommand \enquote  [1]{``#1''}%
\providecommand \bibnamefont  [1]{#1}%
\providecommand \bibfnamefont [1]{#1}%
\providecommand \citenamefont [1]{#1}%
\providecommand \href@noop [0]{\@secondoftwo}%
\providecommand \href [0]{\begingroup \@sanitize@url \@href}%
\providecommand \@href[1]{\@@startlink{#1}\@@href}%
\providecommand \@@href[1]{\endgroup#1\@@endlink}%
\providecommand \@sanitize@url [0]{\catcode `\\12\catcode `\$12\catcode
  `\&12\catcode `\#12\catcode `\^12\catcode `\_12\catcode `\%12\relax}%
\providecommand \@@startlink[1]{}%
\providecommand \@@endlink[0]{}%
\providecommand \url  [0]{\begingroup\@sanitize@url \@url }%
\providecommand \@url [1]{\endgroup\@href {#1}{\urlprefix }}%
\providecommand \urlprefix  [0]{URL }%
\providecommand \Eprint [0]{\href }%
\providecommand \doibase [0]{http://dx.doi.org/}%
\providecommand \selectlanguage [0]{\@gobble}%
\providecommand \bibinfo  [0]{\@secondoftwo}%
\providecommand \bibfield  [0]{\@secondoftwo}%
\providecommand \translation [1]{[#1]}%
\providecommand \BibitemOpen [0]{}%
\providecommand \bibitemStop [0]{}%
\providecommand \bibitemNoStop [0]{.\EOS\space}%
\providecommand \EOS [0]{\spacefactor3000\relax}%
\providecommand \BibitemShut  [1]{\csname bibitem#1\endcsname}%
\let\auto@bib@innerbib\@empty
\bibitem [{\citenamefont {Buchert}\ \emph {et~al.}(2017)\citenamefont
  {Buchert}, \citenamefont {Coley}, \citenamefont {Kleinert}, \citenamefont
  {Roukema},\ and\ \citenamefont {Wiltshire}}]{Buchert:2015wwr}%
  \BibitemOpen
  \bibfield  {author} {\bibinfo {author} {\bibfnamefont {Thomas}\ \bibnamefont
  {Buchert}}, \bibinfo {author} {\bibfnamefont {Alan~A.}\ \bibnamefont
  {Coley}}, \bibinfo {author} {\bibfnamefont {Hagen}\ \bibnamefont {Kleinert}},
  \bibinfo {author} {\bibfnamefont {Boudewijn~F.}\ \bibnamefont {Roukema}}, \
  and\ \bibinfo {author} {\bibfnamefont {David~L.}\ \bibnamefont {Wiltshire}},\
  }\bibfield  {title} {\enquote {\bibinfo {title} {{Observational Challenges
  for the Standard FLRW Model}},}\ }\href {\doibase 10.1142/S021827181630007X}
  {\ \textbf {\bibinfo {volume} {1}},\ \bibinfo {pages} {622--638} (\bibinfo
  {year} {2017})},\ \Eprint {http://arxiv.org/abs/1512.03313} {arXiv:1512.03313
  [astro-ph.CO]} \BibitemShut {NoStop}%
\bibitem [{\citenamefont {Verde}\ \emph {et~al.}(2019)\citenamefont {Verde},
  \citenamefont {Treu},\ and\ \citenamefont {Riess}}]{Verde:2019ivm}%
  \BibitemOpen
  \bibfield  {author} {\bibinfo {author} {\bibfnamefont {L.}~\bibnamefont
  {Verde}}, \bibinfo {author} {\bibfnamefont {T.}~\bibnamefont {Treu}}, \ and\
  \bibinfo {author} {\bibfnamefont {A.G.}\ \bibnamefont {Riess}},\ }\bibfield
  {title} {\enquote {\bibinfo {title} {{Tensions between the Early and the Late
  Universe}},}\ \ }(\bibinfo {year} {2019})\ \Eprint
  {http://arxiv.org/abs/1907.10625} {arXiv:1907.10625 [astro-ph.CO]}
  \BibitemShut {NoStop}%
\bibitem [{\citenamefont {Ellis}\ and\ \citenamefont
  {Stoeger}(1987)}]{Ellis:1987zz}%
  \BibitemOpen
  \bibfield  {author} {\bibinfo {author} {\bibfnamefont {G.F.R.}\ \bibnamefont
  {Ellis}}\ and\ \bibinfo {author} {\bibfnamefont {W.}~\bibnamefont
  {Stoeger}},\ }\bibfield  {title} {\enquote {\bibinfo {title} {{The 'fitting
  problem' in cosmology}},}\ }\href {\doibase 10.1088/0264-9381/4/6/025}
  {\bibfield  {journal} {\bibinfo  {journal} {Class. Quant. Grav.}\ }\textbf
  {\bibinfo {volume} {4}},\ \bibinfo {pages} {1697--1729} (\bibinfo {year}
  {1987})}\BibitemShut {NoStop}%
\bibitem [{\citenamefont {Kolb}\ \emph {et~al.}(2010)\citenamefont {Kolb},
  \citenamefont {Marra},\ and\ \citenamefont {Matarrese}}]{Kolb:2009rp}%
  \BibitemOpen
  \bibfield  {author} {\bibinfo {author} {\bibfnamefont {Edward~W.}\
  \bibnamefont {Kolb}}, \bibinfo {author} {\bibfnamefont {Valerio}\
  \bibnamefont {Marra}}, \ and\ \bibinfo {author} {\bibfnamefont {Sabino}\
  \bibnamefont {Matarrese}},\ }\bibfield  {title} {\enquote {\bibinfo {title}
  {{Cosmological background solutions and cosmological backreactions}},}\
  }\href {\doibase 10.1007/s10714-009-0913-8} {\bibfield  {journal} {\bibinfo
  {journal} {Gen. Rel. Grav.}\ }\textbf {\bibinfo {volume} {42}},\ \bibinfo
  {pages} {1399--1412} (\bibinfo {year} {2010})},\ \Eprint
  {http://arxiv.org/abs/0901.4566} {arXiv:0901.4566 [astro-ph.CO]} \BibitemShut
  {NoStop}%
\bibitem [{\citenamefont {Clarkson}\ \emph {et~al.}(2011)\citenamefont
  {Clarkson}, \citenamefont {Ellis}, \citenamefont {Larena},\ and\
  \citenamefont {Umeh}}]{Clarkson:2011zq}%
  \BibitemOpen
  \bibfield  {author} {\bibinfo {author} {\bibfnamefont {Chris}\ \bibnamefont
  {Clarkson}}, \bibinfo {author} {\bibfnamefont {George}\ \bibnamefont
  {Ellis}}, \bibinfo {author} {\bibfnamefont {Julien}\ \bibnamefont {Larena}},
  \ and\ \bibinfo {author} {\bibfnamefont {Obinna}\ \bibnamefont {Umeh}},\
  }\bibfield  {title} {\enquote {\bibinfo {title} {{Does the growth of
  structure affect our dynamical models of the universe? The averaging,
  backreaction and fitting problems in cosmology}},}\ }\href {\doibase
  10.1088/0034-4885/74/11/112901} {\bibfield  {journal} {\bibinfo  {journal}
  {Rept. Prog. Phys.}\ }\textbf {\bibinfo {volume} {74}},\ \bibinfo {pages}
  {112901} (\bibinfo {year} {2011})},\ \Eprint {http://arxiv.org/abs/1109.2314}
  {arXiv:1109.2314 [astro-ph.CO]} \BibitemShut {NoStop}%
\bibitem [{\citenamefont {Ellis}\ \emph {et~al.}(2012)\citenamefont {Ellis},
  \citenamefont {Maartens},\ and\ \citenamefont
  {MacCallum}}]{ellis_maartens_maccallum_2012}%
  \BibitemOpen
  \bibfield  {author} {\bibinfo {author} {\bibfnamefont {George F.~R.}\
  \bibnamefont {Ellis}}, \bibinfo {author} {\bibfnamefont {Roy}\ \bibnamefont
  {Maartens}}, \ and\ \bibinfo {author} {\bibfnamefont {Malcolm A.~H.}\
  \bibnamefont {MacCallum}},\ }\href {\doibase 10.1017/CBO9781139014403} {\emph
  {\bibinfo {title} {Relativistic Cosmology}}}\ (\bibinfo  {publisher}
  {Cambridge University Press},\ \bibinfo {year} {2012})\BibitemShut {NoStop}%
\bibitem [{\citenamefont {Gasperini}\ \emph {et~al.}(2011)\citenamefont
  {Gasperini}, \citenamefont {Marozzi}, \citenamefont {Nugier},\ and\
  \citenamefont {Veneziano}}]{Gasperini:2011us}%
  \BibitemOpen
  \bibfield  {author} {\bibinfo {author} {\bibfnamefont {M.}~\bibnamefont
  {Gasperini}}, \bibinfo {author} {\bibfnamefont {G.}~\bibnamefont {Marozzi}},
  \bibinfo {author} {\bibfnamefont {F.}~\bibnamefont {Nugier}}, \ and\ \bibinfo
  {author} {\bibfnamefont {G.}~\bibnamefont {Veneziano}},\ }\bibfield  {title}
  {\enquote {\bibinfo {title} {{Light-cone averaging in cosmology: Formalism
  and applications}},}\ }\href {\doibase 10.1088/1475-7516/2011/07/008}
  {\bibfield  {journal} {\bibinfo  {journal} {JCAP}\ }\textbf {\bibinfo
  {volume} {07}},\ \bibinfo {pages} {008} (\bibinfo {year} {2011})},\ \Eprint
  {http://arxiv.org/abs/1104.1167} {arXiv:1104.1167 [astro-ph.CO]} \BibitemShut
  {NoStop}%
\bibitem [{\citenamefont {Fanizza}\ \emph {et~al.}(2020)\citenamefont
  {Fanizza}, \citenamefont {Gasperini}, \citenamefont {Marozzi},\ and\
  \citenamefont {Veneziano}}]{Fanizza:2019pfp}%
  \BibitemOpen
  \bibfield  {author} {\bibinfo {author} {\bibfnamefont {Giuseppe}\
  \bibnamefont {Fanizza}}, \bibinfo {author} {\bibfnamefont {Maurizio}\
  \bibnamefont {Gasperini}}, \bibinfo {author} {\bibfnamefont {Giovanni}\
  \bibnamefont {Marozzi}}, \ and\ \bibinfo {author} {\bibfnamefont {Gabriele}\
  \bibnamefont {Veneziano}},\ }\bibfield  {title} {\enquote {\bibinfo {title}
  {{Generalized covariant prescriptions for averaging cosmological
  observables}},}\ }\href {\doibase 10.1088/1475-7516/2020/02/017} {\bibfield
  {journal} {\bibinfo  {journal} {JCAP}\ }\textbf {\bibinfo {volume} {02}},\
  \bibinfo {pages} {017} (\bibinfo {year} {2020})},\ \Eprint
  {http://arxiv.org/abs/1911.09469} {arXiv:1911.09469 [gr-qc]} \BibitemShut
  {NoStop}%
\bibitem [{\citenamefont {Buchert}(2000)}]{Buchert:1999er}%
  \BibitemOpen
  \bibfield  {author} {\bibinfo {author} {\bibfnamefont {Thomas}\ \bibnamefont
  {Buchert}},\ }\bibfield  {title} {\enquote {\bibinfo {title} {{On average
  properties of inhomogeneous fluids in general relativity. 1. Dust
  cosmologies}},}\ }\href {\doibase 10.1023/A:1001800617177} {\bibfield
  {journal} {\bibinfo  {journal} {Gen. Rel. Grav.}\ }\textbf {\bibinfo {volume}
  {32}},\ \bibinfo {pages} {105--125} (\bibinfo {year} {2000})},\ \Eprint
  {http://arxiv.org/abs/gr-qc/9906015} {arXiv:gr-qc/9906015} \BibitemShut
  {NoStop}%
\bibitem [{\citenamefont {Buchert}(2001)}]{Buchert:2001sa}%
  \BibitemOpen
  \bibfield  {author} {\bibinfo {author} {\bibfnamefont {Thomas}\ \bibnamefont
  {Buchert}},\ }\bibfield  {title} {\enquote {\bibinfo {title} {{On average
  properties of inhomogeneous fluids in general relativity: Perfect fluid
  cosmologies}},}\ }\href {\doibase 10.1023/A:1012061725841} {\bibfield
  {journal} {\bibinfo  {journal} {Gen. Rel. Grav.}\ }\textbf {\bibinfo {volume}
  {33}},\ \bibinfo {pages} {1381--1405} (\bibinfo {year} {2001})},\ \Eprint
  {http://arxiv.org/abs/gr-qc/0102049} {arXiv:gr-qc/0102049} \BibitemShut
  {NoStop}%
\bibitem [{\citenamefont {Bonvin}\ \emph {et~al.}(2015)\citenamefont {Bonvin},
  \citenamefont {Clarkson}, \citenamefont {Durrer}, \citenamefont {Maartens},\
  and\ \citenamefont {Umeh}}]{Bonvin_2015}%
  \BibitemOpen
  \bibfield  {author} {\bibinfo {author} {\bibfnamefont {Camille}\ \bibnamefont
  {Bonvin}}, \bibinfo {author} {\bibfnamefont {Chris}\ \bibnamefont
  {Clarkson}}, \bibinfo {author} {\bibfnamefont {Ruth}\ \bibnamefont {Durrer}},
  \bibinfo {author} {\bibfnamefont {Roy}\ \bibnamefont {Maartens}}, \ and\
  \bibinfo {author} {\bibfnamefont {Obinna}\ \bibnamefont {Umeh}},\ }\bibfield
  {title} {\enquote {\bibinfo {title} {Cosmological ensemble and directional
  averages of observables},}\ }\href {\doibase 10.1088/1475-7516/2015/07/040}
  {\bibfield  {journal} {\bibinfo  {journal} {Journal of Cosmology and
  Astroparticle Physics}\ }\textbf {\bibinfo {volume} {2015}},\ \bibinfo
  {pages} {040--040} (\bibinfo {year} {2015})}\BibitemShut {NoStop}%
\bibitem [{\citenamefont {Buchert}\ and\ \citenamefont
  {Carfora}(2002)}]{Buchert:2002ht}%
  \BibitemOpen
  \bibfield  {author} {\bibinfo {author} {\bibfnamefont {Thomas}\ \bibnamefont
  {Buchert}}\ and\ \bibinfo {author} {\bibfnamefont {Mauro}\ \bibnamefont
  {Carfora}},\ }\bibfield  {title} {\enquote {\bibinfo {title} {{Regional
  averaging and scaling in relativistic cosmology}},}\ }\href {\doibase
  10.1088/0264-9381/19/23/314} {\bibfield  {journal} {\bibinfo  {journal}
  {Class. Quant. Grav.}\ }\textbf {\bibinfo {volume} {19}},\ \bibinfo {pages}
  {6109--6145} (\bibinfo {year} {2002})},\ \Eprint
  {http://arxiv.org/abs/gr-qc/0210037} {arXiv:gr-qc/0210037} \BibitemShut
  {NoStop}%
\bibitem [{\citenamefont {Zalaletdinov}(2008)}]{Zalaletdinov:2008ts}%
  \BibitemOpen
  \bibfield  {author} {\bibinfo {author} {\bibfnamefont {Roustam}\ \bibnamefont
  {Zalaletdinov}},\ }\bibfield  {title} {\enquote {\bibinfo {title} {{The
  Averaging Problem in Cosmology and Macroscopic Gravity}},}\ }\href {\doibase
  10.1142/S0217751X08040032} {\bibfield  {journal} {\bibinfo  {journal} {Int.
  J. Mod. Phys. A}\ }\textbf {\bibinfo {volume} {23}},\ \bibinfo {pages}
  {1173--1181} (\bibinfo {year} {2008})},\ \Eprint
  {http://arxiv.org/abs/0801.3256} {arXiv:0801.3256 [gr-qc]} \BibitemShut
  {NoStop}%
\bibitem [{\citenamefont {Buchert}\ and\ \citenamefont
  {R\"as\"anen}(2012)}]{Buchert:2011sx}%
  \BibitemOpen
  \bibfield  {author} {\bibinfo {author} {\bibfnamefont {Thomas}\ \bibnamefont
  {Buchert}}\ and\ \bibinfo {author} {\bibfnamefont {Syksy}\ \bibnamefont
  {R\"as\"anen}},\ }\bibfield  {title} {\enquote {\bibinfo {title}
  {{Backreaction in late-time cosmology}},}\ }\href {\doibase
  10.1146/annurev.nucl.012809.104435} {\bibfield  {journal} {\bibinfo
  {journal} {Ann. Rev. Nucl. Part. Sci.}\ }\textbf {\bibinfo {volume} {62}},\
  \bibinfo {pages} {57--79} (\bibinfo {year} {2012})},\ \Eprint
  {http://arxiv.org/abs/1112.5335} {arXiv:1112.5335 [astro-ph.CO]} \BibitemShut
  {NoStop}%
\bibitem [{\citenamefont {Green}\ and\ \citenamefont
  {Wald}(2014)}]{Green:2014aga}%
  \BibitemOpen
  \bibfield  {author} {\bibinfo {author} {\bibfnamefont {Stephen~R.}\
  \bibnamefont {Green}}\ and\ \bibinfo {author} {\bibfnamefont {Robert~M.}\
  \bibnamefont {Wald}},\ }\bibfield  {title} {\enquote {\bibinfo {title} {{How
  well is our universe described by an FLRW model?}}}\ }\href {\doibase
  10.1088/0264-9381/31/23/234003} {\bibfield  {journal} {\bibinfo  {journal}
  {Class. Quant. Grav.}\ }\textbf {\bibinfo {volume} {31}},\ \bibinfo {pages}
  {234003} (\bibinfo {year} {2014})},\ \Eprint {http://arxiv.org/abs/1407.8084}
  {arXiv:1407.8084 [gr-qc]} \BibitemShut {NoStop}%
\bibitem [{\citenamefont {Buchert}\ \emph {et~al.}(2015)\citenamefont {Buchert}
  \emph {et~al.}}]{Buchert:2015iva}%
  \BibitemOpen
  \bibfield  {author} {\bibinfo {author} {\bibfnamefont {T.}~\bibnamefont
  {Buchert}} \emph {et~al.},\ }\bibfield  {title} {\enquote {\bibinfo {title}
  {{Is there proof that backreaction of inhomogeneities is irrelevant in
  cosmology?}}}\ }\href {\doibase 10.1088/0264-9381/32/21/215021} {\bibfield
  {journal} {\bibinfo  {journal} {Class. Quant. Grav.}\ }\textbf {\bibinfo
  {volume} {32}},\ \bibinfo {pages} {215021} (\bibinfo {year} {2015})},\
  \Eprint {http://arxiv.org/abs/1505.07800} {arXiv:1505.07800 [gr-qc]}
  \BibitemShut {NoStop}%
\bibitem [{\citenamefont {Larena}\ \emph {et~al.}(2009)\citenamefont {Larena},
  \citenamefont {Alimi}, \citenamefont {Buchert}, \citenamefont {Kunz},\ and\
  \citenamefont {Corasaniti}}]{Larena:2008be}%
  \BibitemOpen
  \bibfield  {author} {\bibinfo {author} {\bibfnamefont {Julien}\ \bibnamefont
  {Larena}}, \bibinfo {author} {\bibfnamefont {Jean-Michel}\ \bibnamefont
  {Alimi}}, \bibinfo {author} {\bibfnamefont {Thomas}\ \bibnamefont {Buchert}},
  \bibinfo {author} {\bibfnamefont {Martin}\ \bibnamefont {Kunz}}, \ and\
  \bibinfo {author} {\bibfnamefont {Pier-Stefano}\ \bibnamefont {Corasaniti}},\
  }\bibfield  {title} {\enquote {\bibinfo {title} {{Testing backreaction
  effects with observations}},}\ }\href {\doibase 10.1103/PhysRevD.79.083011}
  {\bibfield  {journal} {\bibinfo  {journal} {Phys. Rev. D}\ }\textbf {\bibinfo
  {volume} {79}},\ \bibinfo {pages} {083011} (\bibinfo {year} {2009})},\
  \Eprint {http://arxiv.org/abs/0808.1161} {arXiv:0808.1161 [astro-ph]}
  \BibitemShut {NoStop}%
\bibitem [{\citenamefont {Ellis}\ \emph {et~al.}(1998)\citenamefont {Ellis},
  \citenamefont {Bassett},\ and\ \citenamefont {Dunsby}}]{Ellis:1998ha}%
  \BibitemOpen
  \bibfield  {author} {\bibinfo {author} {\bibfnamefont {G.F.R.}\ \bibnamefont
  {Ellis}}, \bibinfo {author} {\bibfnamefont {B.A.}\ \bibnamefont {Bassett}}, \
  and\ \bibinfo {author} {\bibfnamefont {P.K.S.}\ \bibnamefont {Dunsby}},\
  }\bibfield  {title} {\enquote {\bibinfo {title} {{Lensing and caustic effects
  on cosmological distances}},}\ }\href {\doibase 10.1088/0264-9381/15/8/015}
  {\bibfield  {journal} {\bibinfo  {journal} {Class. Quant. Grav.}\ }\textbf
  {\bibinfo {volume} {15}},\ \bibinfo {pages} {2345--2361} (\bibinfo {year}
  {1998})},\ \Eprint {http://arxiv.org/abs/gr-qc/9801092} {arXiv:gr-qc/9801092}
  \BibitemShut {NoStop}%
\bibitem [{\citenamefont {Hawking}(1968)}]{Hawking:1968qt}%
  \BibitemOpen
  \bibfield  {author} {\bibinfo {author} {\bibfnamefont {Stephen}\ \bibnamefont
  {Hawking}},\ }\bibfield  {title} {\enquote {\bibinfo {title} {{Gravitational
  radiation in an expanding universe}},}\ }\href {\doibase 10.1063/1.1664615}
  {\bibfield  {journal} {\bibinfo  {journal} {J. Math. Phys.}\ }\textbf
  {\bibinfo {volume} {9}},\ \bibinfo {pages} {598--604} (\bibinfo {year}
  {1968})}\BibitemShut {NoStop}%
\bibitem [{\citenamefont {Szabados}(2009)}]{Szabados:2009eka}%
  \BibitemOpen
  \bibfield  {author} {\bibinfo {author} {\bibfnamefont {L\'aszl\'o~B.}\
  \bibnamefont {Szabados}},\ }\bibfield  {title} {\enquote {\bibinfo {title}
  {{Quasi-Local Energy-Momentum and Angular Momentum in General Relativity}},}\
  }\href {\doibase 10.12942/lrr-2009-4} {\bibfield  {journal} {\bibinfo
  {journal} {Living Rev. Rel.}\ }\textbf {\bibinfo {volume} {12}},\ \bibinfo
  {pages} {4} (\bibinfo {year} {2009})}\BibitemShut {NoStop}%
\bibitem [{\citenamefont {Stock}(2020)}]{Stock:2020oda}%
  \BibitemOpen
  \bibfield  {author} {\bibinfo {author} {\bibfnamefont {Dennis}\ \bibnamefont
  {Stock}},\ }\bibfield  {title} {\enquote {\bibinfo {title} {{The Hawking
  Energy on the Past Lightcone in Cosmology}},}\ }\href {\doibase
  10.1088/1361-6382/aba182} {\bibfield  {journal} {\bibinfo  {journal} {Class.
  Quant. Grav.}\ }\textbf {\bibinfo {volume} {37}},\ \bibinfo {pages} {215005}
  (\bibinfo {year} {2020})},\ \Eprint {http://arxiv.org/abs/2003.13583}
  {arXiv:2003.13583 [gr-qc]} \BibitemShut {NoStop}%
\bibitem [{\citenamefont {Ehlers}(1993)}]{Ehlers:1993gf}%
  \BibitemOpen
  \bibfield  {author} {\bibinfo {author} {\bibfnamefont {J.}~\bibnamefont
  {Ehlers}},\ }\bibfield  {title} {\enquote {\bibinfo {title} {{Contributions
  to the relativistic mechanics of continuous media}},}\ }\href {\doibase
  10.1007/BF00759031} {\bibfield  {journal} {\bibinfo  {journal} {Gen. Rel.
  Grav.}\ }\textbf {\bibinfo {volume} {25}},\ \bibinfo {pages} {1225--1266}
  (\bibinfo {year} {1993})}\BibitemShut {NoStop}%
\bibitem [{\citenamefont {Gourgoulhon}\ and\ \citenamefont
  {Jaramillo}(2006)}]{Gourgoulhon:2005ng}%
  \BibitemOpen
  \bibfield  {author} {\bibinfo {author} {\bibfnamefont {Eric}\ \bibnamefont
  {Gourgoulhon}}\ and\ \bibinfo {author} {\bibfnamefont {Jose~Luis}\
  \bibnamefont {Jaramillo}},\ }\bibfield  {title} {\enquote {\bibinfo {title}
  {{A 3+1 perspective on null hypersurfaces and isolated horizons}},}\ }\href
  {\doibase 10.1016/j.physrep.2005.10.005} {\bibfield  {journal} {\bibinfo
  {journal} {Phys. Rept.}\ }\textbf {\bibinfo {volume} {423}},\ \bibinfo
  {pages} {159--294} (\bibinfo {year} {2006})},\ \Eprint
  {http://arxiv.org/abs/gr-qc/0503113} {arXiv:gr-qc/0503113} \BibitemShut
  {NoStop}%
\bibitem [{\citenamefont {Wald}(1984)}]{Wald:1984rg}%
  \BibitemOpen
  \bibfield  {author} {\bibinfo {author} {\bibfnamefont {Robert~M.}\
  \bibnamefont {Wald}},\ }\href {\doibase
  10.7208/chicago/9780226870373.001.0001} {\emph {\bibinfo {title} {{General
  Relativity}}}}\ (\bibinfo  {publisher} {Chicago Univ. Pr.},\ \bibinfo
  {address} {Chicago, USA},\ \bibinfo {year} {1984})\BibitemShut {NoStop}%
\bibitem [{\citenamefont {Hawking}\ and\ \citenamefont
  {Ellis}(2011)}]{Hawking:1973uf}%
  \BibitemOpen
  \bibfield  {author} {\bibinfo {author} {\bibfnamefont {S.W.}\ \bibnamefont
  {Hawking}}\ and\ \bibinfo {author} {\bibfnamefont {G.F.R.}\ \bibnamefont
  {Ellis}},\ }\href {\doibase 10.1017/CBO9780511524646} {\emph {\bibinfo
  {title} {{The Large Scale Structure of Space-Time}}}},\ Cambridge Monographs
  on Mathematical Physics\ (\bibinfo  {publisher} {Cambridge University
  Press},\ \bibinfo {year} {2011})\BibitemShut {NoStop}%
\bibitem [{\citenamefont {Gibbons}\ and\ \citenamefont
  {Solodukhin}(2007)}]{Gibbons:2007nm}%
  \BibitemOpen
  \bibfield  {author} {\bibinfo {author} {\bibfnamefont {G.W.}\ \bibnamefont
  {Gibbons}}\ and\ \bibinfo {author} {\bibfnamefont {S.N.}\ \bibnamefont
  {Solodukhin}},\ }\bibfield  {title} {\enquote {\bibinfo {title} {{The
  Geometry of small causal diamonds}},}\ }\href {\doibase
  10.1016/j.physletb.2007.03.068} {\bibfield  {journal} {\bibinfo  {journal}
  {Phys. Lett. B}\ }\textbf {\bibinfo {volume} {649}},\ \bibinfo {pages}
  {317--324} (\bibinfo {year} {2007})},\ \Eprint
  {http://arxiv.org/abs/hep-th/0703098} {arXiv:hep-th/0703098} \BibitemShut
  {NoStop}%
\bibitem [{\citenamefont {Wang}(2019)}]{Wang:2019zhr}%
  \BibitemOpen
  \bibfield  {author} {\bibinfo {author} {\bibfnamefont {Jinzhao}\ \bibnamefont
  {Wang}},\ }\bibfield  {title} {\enquote {\bibinfo {title} {{Geometry of small
  causal diamonds}},}\ }\href {\doibase 10.1103/PhysRevD.100.064020} {\bibfield
   {journal} {\bibinfo  {journal} {Phys. Rev. D}\ }\textbf {\bibinfo {volume}
  {100}},\ \bibinfo {pages} {064020} (\bibinfo {year} {2019})},\ \Eprint
  {http://arxiv.org/abs/1904.01034} {arXiv:1904.01034 [gr-qc]} \BibitemShut
  {NoStop}%
\bibitem [{\citenamefont {Choquet-Bruhat}\ \emph {et~al.}(2009)\citenamefont
  {Choquet-Bruhat}, \citenamefont {Chrusciel},\ and\ \citenamefont
  {Martin-Garcia}}]{ChoquetBruhat:2009fy}%
  \BibitemOpen
  \bibfield  {author} {\bibinfo {author} {\bibfnamefont {Yvonne}\ \bibnamefont
  {Choquet-Bruhat}}, \bibinfo {author} {\bibfnamefont {Piotr~T.}\ \bibnamefont
  {Chrusciel}}, \ and\ \bibinfo {author} {\bibfnamefont {Jose~M.}\ \bibnamefont
  {Martin-Garcia}},\ }\bibfield  {title} {\enquote {\bibinfo {title} {{The
  Light-cone theorem}},}\ }\href {\doibase 10.1088/0264-9381/26/13/135011}
  {\bibfield  {journal} {\bibinfo  {journal} {Class. Quant. Grav.}\ }\textbf
  {\bibinfo {volume} {26}},\ \bibinfo {pages} {135011} (\bibinfo {year}
  {2009})},\ \Eprint {http://arxiv.org/abs/0905.2133} {arXiv:0905.2133 [gr-qc]}
  \BibitemShut {NoStop}%
\bibitem [{\citenamefont {Horowitz}\ and\ \citenamefont
  {Schmidt}(1982)}]{Horowitz:1982}%
  \BibitemOpen
  \bibfield  {author} {\bibinfo {author} {\bibfnamefont {G.~T.}\ \bibnamefont
  {Horowitz}}\ and\ \bibinfo {author} {\bibfnamefont {B.~G.}\ \bibnamefont
  {Schmidt}},\ }\bibfield  {title} {\enquote {\bibinfo {title} {Note on
  gravitational energy},}\ }\href {http://www.jstor.org/stable/2397373}
  {\bibfield  {journal} {\bibinfo  {journal} {Proceedings of the Royal Society
  of London. Series A, Mathematical and Physical Sciences}\ }\textbf {\bibinfo
  {volume} {381}},\ \bibinfo {pages} {215--224} (\bibinfo {year}
  {1982})}\BibitemShut {NoStop}%
\bibitem [{\citenamefont {{Planck Collaboration}}(2020)}]{Aghanim:2018eyx}%
  \BibitemOpen
  \bibfield  {author} {\bibinfo {author} {\bibnamefont {{Planck
  Collaboration}}},\ }\bibfield  {title} {\enquote {\bibinfo {title} {{Planck
  2018 results. VI. Cosmological parameters}},}\ }\href {\doibase
  10.1051/0004-6361/201833910} {\bibfield  {journal} {\bibinfo  {journal}
  {Astron. Astrophys.}\ }\textbf {\bibinfo {volume} {641}},\ \bibinfo {pages}
  {A6} (\bibinfo {year} {2020})},\ \Eprint {http://arxiv.org/abs/1807.06209}
  {arXiv:1807.06209 [astro-ph.CO]} \BibitemShut {NoStop}%
\end{thebibliography}%

\end{document}